\documentclass[aps,pra,a4paper,showpacs,twocolumn,floatfix]{revtex4}
\usepackage[latin1]{inputenc}
\usepackage{amsfonts}
\usepackage{amssymb}
\usepackage{amsmath}
\usepackage{calc}
\usepackage{graphicx}
\usepackage{epsfig}
\usepackage{bm}
\usepackage{ulem}
\usepackage{setspace}
\usepackage{subfigure}

\begin{document}

\title{Robust and Efficient Quantum Repeaters with Atomic Ensembles and Linear Optics}
\date{\today}
\pacs{03.67.Hk, 03.67.Mn, 42.50.Md, 76.30.Kg}
\author{Nicolas Sangouard,$^{1}$  Christoph Simon,$^{2}$
Bo Zhao,$^{3}$ Yu-Ao Chen,$^{3}$ Hugues de Riedmatten,$^2$
Jian-Wei Pan$^{3}$, and Nicolas Gisin$^{2}$}
\affiliation{%
$^{1}$Laboratoire MPQ, UMR CNRS 7162, Universit\'e Paris 7, France\\
$^{2}$Group of Applied Physics, University of Geneva, Switzerland\\
$^3$ Physikalisches Institut, Universit\"at Heidelberg,
Germany}

\begin{abstract}
In the last few years there has been a lot of interest in
quantum repeater protocols using only atomic ensembles and
linear optics. Here we show that the local generation of
high-fidelity entangled pairs of atomic excitations, in
combination with the use of two-photon detections for
long-distance entanglement generation, permits the
implementation of a very attractive quantum repeater
protocol. Such a repeater is robust with respect to phase
fluctuations in the transmission channels, and at the same
time achieves higher entanglement generation rates than
other protocols using the same ingredients. We propose an
efficient method of generating high-fidelity entangled
pairs locally, based on the partial readout of the
ensemble-based memories. We also discuss the experimental
implementation of the proposed protocol.
\end{abstract}

\maketitle

\section{Introduction}

The distribution of entangled states over long distances is
difficult because of unavoidable transmission losses and
the no-cloning theorem for quantum states. One possible
solution is the use of quantum repeaters \cite{Briegel98}.
In this approach, entanglement is generated independently
for relatively short elementary links and stored in quantum
memories. Entanglement over longer distances can then be
created by entanglement swapping.

The Duan-Lukin-Cirac-Zoller (DLCZ) protocol \cite{Duan01}
relies on Raman scattering in atomic ensembles, which can
create a single ``Stokes'' photon correlated with a
collective atomic excitation. Two remote atomic ensembles
can then be entangled based on the detection at a central
station of a single Stokes photon, which could have been
emitted by either of the two ensembles \cite{Chou05}. The
entanglement can then be extended to long distances by
converting back the atomic excitations into ``Anti-Stokes''
photons via the reverse Raman process and performing
successive entanglement swapping operations, which are also
based on the detection of a single Anti-Stokes photon.

The DLCZ protocol is attractive because it uses relatively
simple ingredients. Over the last few years there has been
a lot of experimental activity towards its realization
\cite{dlcz_expts}, including the creation of entanglement
between separate quantum nodes \cite{chou} and the
realization of teleportation between photonic and atomic
qubits \cite{yachen}. Conversion efficiencies from atomic
to photonic excitations as high as 84 percent have recently
been achieved for ensembles inside optical cavities
\cite{simon-vuletic}.

However the DLCZ protocol also has a certain number of
practical drawbacks. On the one hand, the {\it generation}
of entanglement via single-photon detections requires
interferometric stability over the whole distance, which a
priori seems quite challenging. For recent experimental
work towards assessing the feasibility of this requirement
for optical fiber links see Ref. \cite{Minar07}.

On the other hand, the {\it swapping} of entanglement using
single-photon detections leads to the growth of a vacuum
component in the generated state, and to the rapid
(quadratic with the number of links) growth of errors due
to multiple emissions from individual ensembles. In order
to suppress these errors, one then has to work with very
low emission probabilities. These factors together lead to
rather low entanglement distribution rates for the DLCZ
protocol. See Fig. 1, curve B, for its performance in the
distance range from 400 km to 1200 km. Moreover, the DLCZ
protocol does not contain a procedure for entanglement
purification (of phase errors in particular), which limits
the total number of links that can be used.

This paper is organized as follows. In section
\ref{Comparison} we compare the entanglement generation
times for the DLCZ protocol and for a number of recently
proposed improved protocols that use the same ingredients
\cite{Jiang07,Sangouard07,Zhao07,Chen07}. We show that an
approach that combines the {\it local generation of
high-fidelity entangled pairs of atomic excitations} and
the creation and swapping of long-distance entanglement via
{\it two-photon detections} is particularly promising in
terms of robustness and achievable entanglement generation
rate. The main drawback  of the implementation of this
approach proposed in Ref. \cite{Chen07} is its relatively
high complexity for local pair generation, which also has a
negative impact on the fidelity of the created pairs for
non-unit memory and detection efficiency. In section
\ref{Pairs} we propose an improved method for the local
generation of entangled pairs of atomic excitations based
on the {\it partial readout} of the atomic ensemble
memories. In section \ref{Repeater} we calculate the
entanglement generation times for a quantum repeater using
this new source of pairs. The resulting protocol is both
robust with respect to phase fluctuations in the
transmission channels and significantly more efficient than
all other protocols (known to us) that use the same
ingredients. In section \ref{Implementation} we discuss the
prospects of experimental implementation.

\begin{figure}[hr!]
{\includegraphics[scale=0.32]{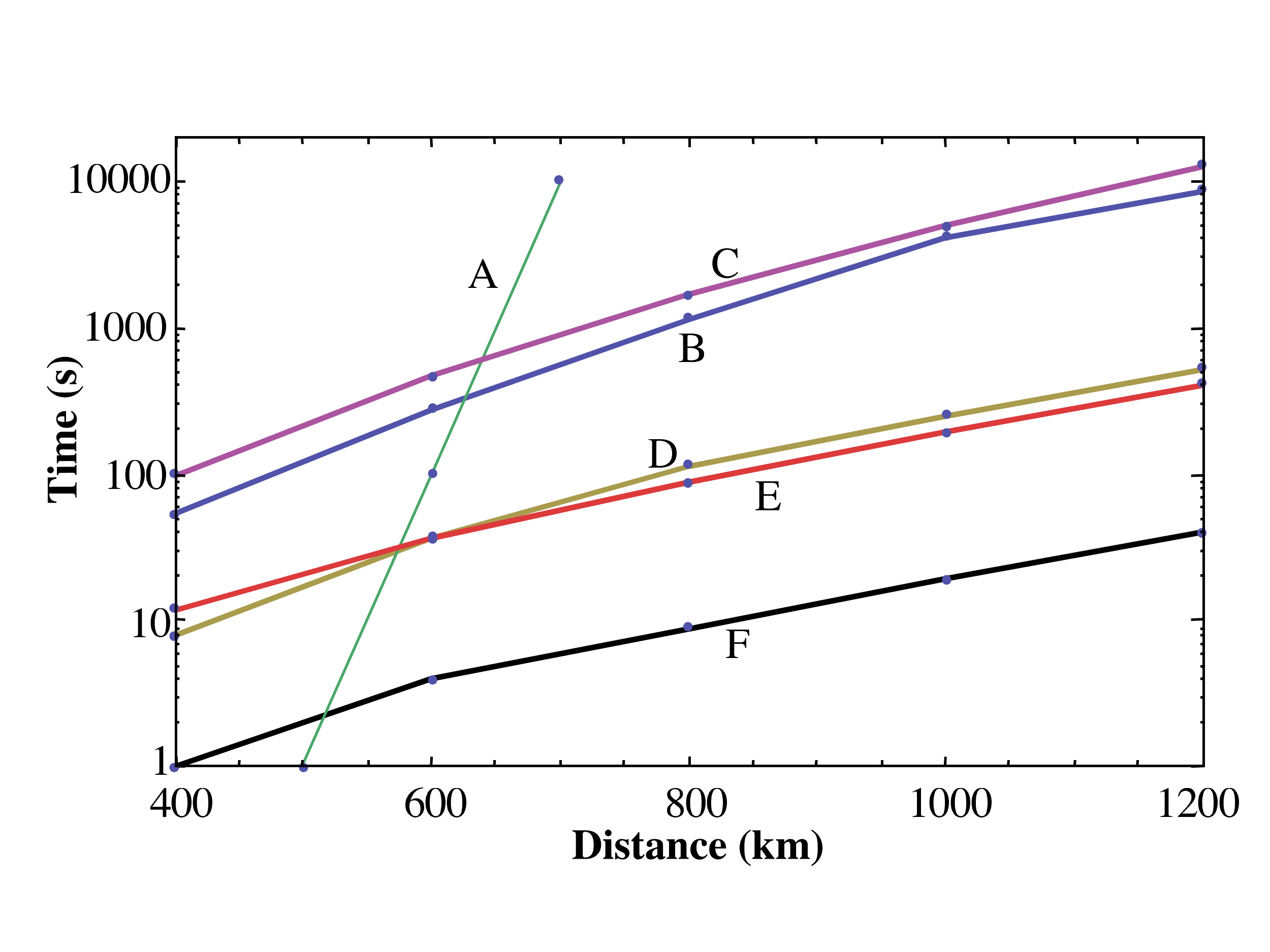} \caption{(Color
online) Comparison of different quantum repeater protocols
that all use only atomic ensembles and linear optics. The
quantity shown is the average time needed to distribute a
single entangled pair for the given distance. A: as a
reference, the time required using direct transmission of
photons through optical fibers, with losses of 0.2 dB/km,
corresponding to the best available telecom fibers at a
wavelength of 1.5 $\mu$m, and a pair generation rate of 10
GHz. B: the original DLCZ protocol that uses single-photon
detections for both entanglement generation and swapping
\cite{Duan01}. C: the protocol of Ref. \cite{Chen07}
section III.B which first creates entanglement locally
using single-photon detections, and then generates
long-distance entanglement using two-photon detections. D:
protocol of Ref. \cite{Sangouard07} that uses quasi-ideal
single photon sources (which can be implemented with atomic
ensembles, cf. text) plus single-photon detections for
generation and swapping. E: protocol of Ref. \cite{Chen07}
section III.C that locally generates high-fidelity
entangled pairs and uses two-photon detections for
entanglement generation and swapping. F: the proposed new
protocol which follows the approach of Ref. \cite{Chen07}
section III.C, but uses an improved method of generating
the local entanglement. The performance of the protocol of
Ref. \cite{Jiang07} is close to curve B for this distance
range (the authors announce a factor of 2 improvement over
DLCZ at 640 km and a factor of 5 at 1280 km). For all the
curves we have assumed memory and detector efficiencies of
90\%. The numbers of links in the repeater chain are
optimized for curves B and D, e.g. giving 4 links for 600
km and 8 links for 1000 km for both protocols. For curves
C, E and F, we imposed a maximum number of 16 links (cf.
text), which is used for all distances for curve C and for
distances greater than 400 km for curves E and
F.}\label{fig1}}
\end{figure}

\section{Modifications of the DLCZ protocol - Comparison}
\label{Comparison}

Ref. \cite{Jiang07} recently proposed a modification of the
DLCZ protocol in which entanglement is still generated by
single-photon detections, but entanglement swapping is
based on two-photon detections. As a consequence, the
vacuum component remains constant under entanglement
swapping, multi-photon errors grow only linearly, and
entanglement purification with linear optics \cite{Pan01}
is possible. However the achieved rates are only slightly
better than for the DLCZ protocol for distances of order
1000 km (see also caption of Fig. 1), mainly because errors
in the elementary link due to multiple excitations still
force one to work with low emission probabilities. Multiple
excitations are hard to detect in the entanglement
generation process because the corresponding Stokes photons
have to propagate far and are lost with high probability.

\begin{table}
\begin{ruledtabular}
\begin{tabular}{cccccccccccc}
\multicolumn{6}{c} {Reference}  & {generation} & {swapping} & {resources} & {links} & {time}  \\ \hline
\multicolumn{6}{c} {\cite{Duan01}} & 1 & 1 & {4} & 8 & B  \\ \hline
\multicolumn{6}{c} {\cite{Jiang07}} & 1 & 2 & {4} & 16 & -  \\ \hline
\multicolumn{6}{c}  {\cite{Chen07} III.B} & 2 & 2 & {8} &16 & C  \\ \hline
\multicolumn{6}{c}  {\cite{Sangouard07}} & 1 & 1  & {4} & 8 & D \\ \hline
\multicolumn{6}{c}  {\cite{Zhao07}} & 2 & 2 & {4} & 16 & -  \\ \hline
\multicolumn{6}{c}  {\cite{Chen07} III.C} & 2 & 2 & {12} & 16 & E  \\ \hline
\multicolumn{6}{c}  {new} & 2 & 2 & 8 & 16 & F
\end{tabular}
\end{ruledtabular}
\caption{Characteristics of the main protocols for quantum
repeaters based on atomic ensembles. Column 1 gives the
reference for the considered protocol. Columns 2 and 3 show
the number of photons detected in the long-distance
entanglement generation step and in the subsequent
entanglement swapping steps respectively. Column 4 gives
the number of atomic ensembles used within each elementary
link. Column 5 gives the number of links for a distance of
1000 km. The number of links is optimized for protocols
\cite{Duan01, Sangouard07} and limited to 16 for the other
protocols, cf. text. For the protocol of ref.
\cite{Jiang07}, we have taken the number of links announced
by the authors for a distance of 640km. The last column
refers to the curves of Fig. \ref{fig1}, where the
performance of the protocols is compared.\label{table1}}
\end{table}

In Ref. \cite{Chen07}, several protocols based on
two-photon detections were presented. The one presented in
section III.B is a simple variation of the proposal of Ref.
\cite{Jiang07}, in which the entanglement generation step
of Ref. \cite{Jiang07} is performed locally and the first
entanglement swapping step of Ref. \cite{Jiang07} is
performed remotely. This protocol does not require
interferometric stability over long distances. However,
excess photon emissions in the remote swapping step remain
undetected due to fiber losses. As a consequence, the
distributed state after the remote swapping has large
vacuum and single-photon components, which lead to small
success probabilities for the subsequent swapping steps,
and thus to a rather low overall entanglement distribution
rate, comparable to the DLCZ protocol (see curve C in
Fig.1).

Ref. \cite{Sangouard07} uses single-photon detections for
entanglement generation and swapping, but the method of
entanglement generation is different with respect to the
DLCZ protocol, relying on single-photon sources. This makes
it possible to improve the distribution rate of entangled
states, thanks to the suppression of multi-photon errors.
This protocol can be realized with atomic ensembles and
linear optics because a quasi-ideal single-photon source
can be constructed based on atomic ensembles of the DLCZ
type \cite{sp-sources}. The probabilistic emission of the
Stokes photon heralds the creation of an atomic excitation
in the ensemble. The charged memory can now be used as a
single-photon source by reconverting the stored excitation
into an Anti-Stokes photon. The probability for this source
to emit two Anti-Stokes photons can be made arbitrarily
small by working with a small emission probability for the
Stokes photon. The price to pay is that the preparation of
the source requires many attempts until the Stokes photon
is emitted. However, these attempts are purely local and
can thus be repeated very fast. The protocol of Ref.
\cite{Sangouard07} is faster than the DLCZ protocol, see
curve D in Fig. 1. However it shares the need for phase
stability, the amplification of vacuum and multi-photon
components, and the absence of a known entanglement
purification procedure.

Ref. \cite{Zhao07} proposed a scheme in which both
entanglement creation and swapping are based on two-photon
detections \cite{twophoton}. In addition to the advantages
mentioned for Ref. \cite{Jiang07}, this protocol no longer
requires interferometric stability over long distances.
However, in the scheme of Ref. \cite{Zhao07}, entanglement
is directly generated over long distances. Since only a
small excitation probability can be used for each
entanglement generation attempt (in order to avoid
multi-photon errors), and since after every attempt one has
to communicate its success or failure over a long distance,
the required entanglement generation time becomes longer
than for the DLCZ protocol.

Ref. \cite{Chen07} proposed a number of protocols without
making a quantitative comparison. We have found that the
best performing of these protocols is the one presented in
section III.C of Ref. \cite{Chen07}. In this approach one
first locally generates high-fidelity entangled pairs of
atomic excitations that are stored in nearby ensembles.
Then long-distance entanglement is generated and swapped
via two-photon detections. We have calculated the
entanglement generation times for this protocol. Its
performance is shown as curve E in Fig. 1. One can see that
the entanglement distribution time is comparable with the
protocol of Ref. \cite{Sangouard07}. Since high-fidelity
pairs are first generated locally, the effective
``excitation probability'' for every long-distance
entanglement generation attempt is essentially equal to
one. Moreover, the vacuum component remains unchanged under
entanglement swapping thanks to the use of two-photon
detections, leading to high success probabilities for the
higher-order swapping operations. To fully profit from this
scheme, the local pair generation rate has to be
sufficiently high, cf. below.

The use of two-photon detections for long-distance
entanglement generation makes the scheme robust with
respect to phase fluctuations in the channels, but also
more sensitive to photon losses than the schemes that use
single-photon detections for the same purpose
\cite{Duan01,Jiang07,Sangouard07}. As a consequence, the
two-photon protocol favors larger numbers of elementary
links for the same distance compared to the single-photon
schemes. In fact, the optimal number of links for 1000 km,
e.g., would be 32. In Fig. 1, we have limited the maximum
number of links used to 16, to keep it more comparable to
the link numbers for the single-photon schemes, and to have
link numbers for which it is plausible that entanglement
purification may not be necessary. Note however that the
scheme is perfectly compatible with the use of linear
optics entanglement purification as proposed in Ref.
\cite{Pan01}, whereas no entanglement purification protocol
is currently known for the schemes of Refs.
\cite{Duan01,Sangouard07}. Furthermore the stability
requirements are much more stringent for the single-photon
schemes, so achieving sufficiently low error rates for a
single-photon based quantum repeater with 8 links is likely
to be more difficult than for a two-photon based quantum
repeater with 16 links. Our comparison method is thus quite
conservative concerning the advantage of the two-photon
protocol.

A related important question is that of complexity, in
particular the number of atomic ensembles required for a
given repeater, compared to the achieved improvement in the
entanglement generation time, see also Table I. The schemes
of Refs. \cite{Duan01,Jiang07,Sangouard07} require just
four ensembles per elementary link \cite{partial-sps}. The
scheme of Ref. \cite{Chen07} section III.C requires twelve,
since every local pair generation uses four single-photon
sources (which can be realized with ensembles as described
above) and two EIT-based ensemble memories
\cite{EIT-memories}. Together with the increased number of
links discussed above, this means that this protocol is
much less efficient than that of Ref. \cite{Sangouard07}.
It should be noted however that the two-photon protocol
\cite{Chen07} section III.C still remains much more robust
with respect to channel phase fluctuations than the
single-photon protocol \cite{Sangouard07}.

In the following we propose a new method for the local
generation of high-fidelity entangled pairs of atomic
excitations. The method uses the available resources
(atomic ensembles) more efficiently, which makes it
possible to achieve {\it higher-fidelity} entanglement for
the same values of memory and detection efficiency, leading
to a significant improvement in the achievable entanglement
generation rate over long distances. The performance of the
improved protocol is shown as curve F in Fig. 1. For the
new protocol, which requires eight memories per elementary
link, the gain in time clearly outweighs the modest
increase in complexity compared to the fastest
single-photon protocol \cite{Sangouard07}. For example, for
1000 km the new protocol uses four times as many memories,
but it is about 13 times faster. (The rate improvement
compared to the DLCZ protocol, which uses the same number
of memories as Ref. \cite{Sangouard07}, is by a factor of
200.) It is thus not only robust, but also the most
efficient repeater protocol known to us for the given
ingredients.

\section{Local generation of high-fidelity entangled pairs based
on partial memory readout} \label{Pairs}

Ref. \cite{Chen07} proposed to generate high-fidelity
entangled pairs of atomic excitations locally by using four
single-photon sources (which can be realized with DLCZ-type
ensembles, cf. above), linear optical elements, and two
EIT-based quantum memories, cf. Fig. 11 of Ref.
\cite{Chen07}. Four photons are emitted by the ensembles
serving as sources, two of them are detected, two are
absorbed again by the EIT memories. This double use of the
memories (emission followed by storage) leads to relatively
large errors (vacuum and single-photon contributions) in
the created state if the memory efficiencies are smaller
than one. These errors then have a negative impact on the
success probabilities of the entanglement generation and
swapping operations, and thus on the overall time needed
for long-distance entanglement distribution.

Here we propose a different method for the local generation
of high-fidelity entangled pairs of atomic excitations,
which is based on the {\it partial readout} of ensemble
memories. Our scheme does not use any emission followed by
storage. For the same memory and detection efficiency, it
leads to higher quality entangled pairs compared to the
method of Ref. \cite{Chen07}, and as a consequence to a
significantly improved rate for the overall quantum
repeater protocol (curve E in Fig. 1). We now describe the
proposed method for local entanglement generation in
detail.

\begin{figure}
{\includegraphics[scale=0.35]{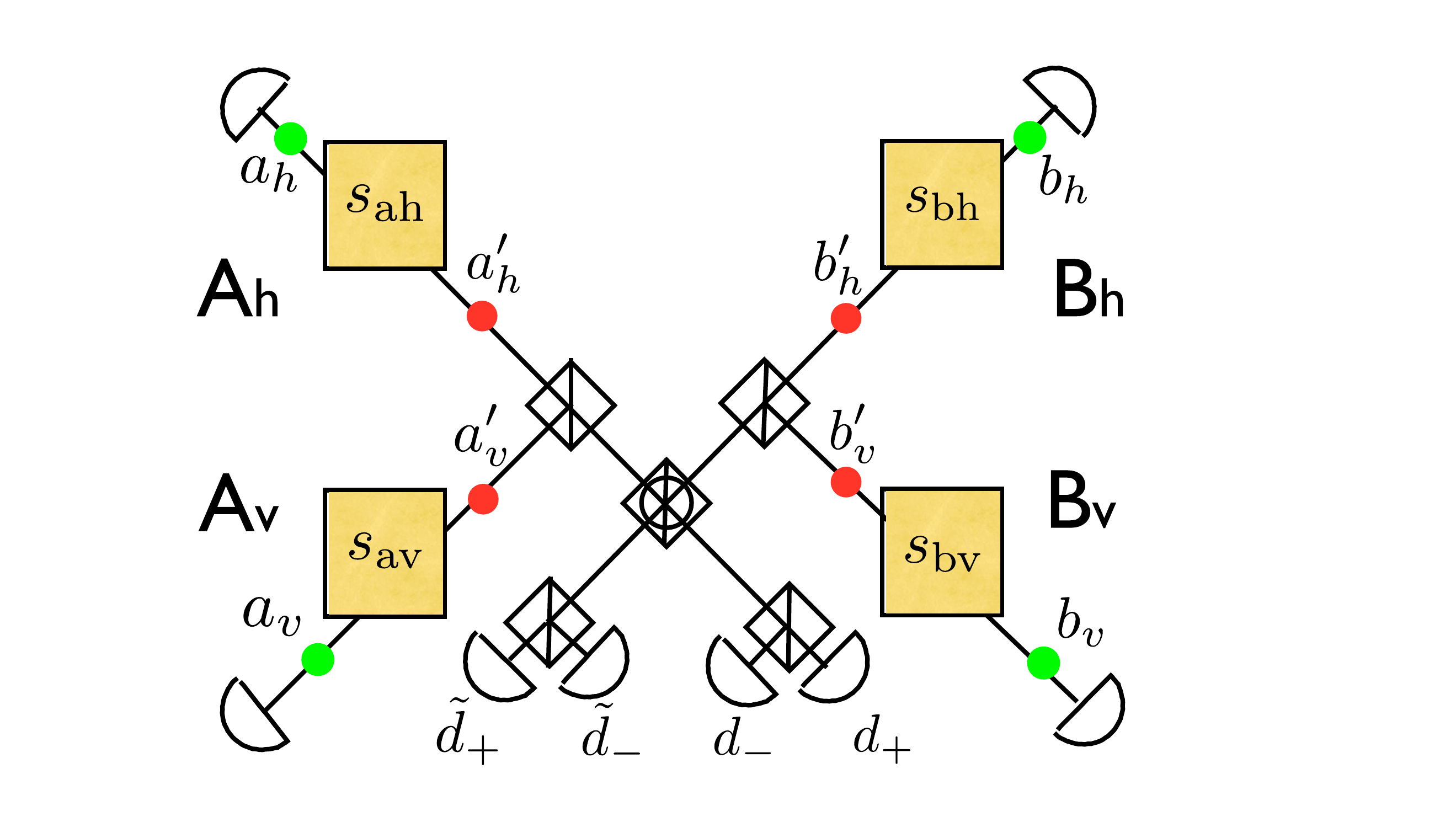} \caption{(Color
online) Setup for generating high-fidelity entangled pairs
of atomic excitations. Yellow squares represent atomic
ensembles which probabilistically emit Stokes photons
(green dots). The conditional detection of a single Stokes
photon heralds the storage of one atomic spin-wave
excitation. In this way an atomic excitation is created and
stored independently in each ensemble. Then all four
ensembles are simultaneously read out {\it partially},
creating a probability amplitude to emit an Anti-Stokes
photon (red dots). The coincident detection of two photons
in ${d}_+$ and ${\tilde{d}}_+$ projects non-destructively
the atomic cells into the entangled state $|\Phi_{\rm
ab}\rangle$ of Eq. (1); ${d}_+$-${\tilde{d}}_-$,
${d}_-$-${\tilde{d}}_+$, and ${d}_-$-${\tilde{d}}_-$
coincidences, combined with the appropriate one-qubit
transformations, also collapse the state of the atomic
cells into $|\Phi_{\rm{ab}}\rangle.$ Half-circles represent
photon detectors. Vertical bars within squares label
polarizing beam splitters (PBS) that transmit (reflect) $H$
($V$)-polarized photons. The central PBS with a circle
performs the same action in the $\pm$ 45$^o$ ($H+V/H-V$)
basis.}\label{fig2}}
\end{figure}

The proposed setup uses four atomic ensembles. Atomic Raman
transitions are coherently excited such that a Stokes
photon can be emitted with a small probability $p$. This
Stokes photon has a well defined polarization : the
horizontally (vertically) polarized modes are labeled by
${a}^{\dagger}_h$ and ${b}^{\dagger}_h,$ $({a}^{\dagger}_v$
and ${b}^{\dagger}_v)$ and are produced from upper (lower)
atomic ensembles $A_h$ and $B_h$ ($A_v$ and $B_v$) as
represented in Fig. \ref{fig2}. The four atomic ensembles
are repeatedly excited \textit{independently} with a
repetition rate $r$ until a Stokes photon has been detected
in each mode ${a}^{\dagger}_h, {a}^{\dagger}_v,
{b}^{\dagger}_h,$ and ${b}^{\dagger}_v.$ The detection of a
Stokes photon heralds the storage of a single atomic spin
excitation in each ensemble, labeled by ${s}^{\dagger}_{\rm
ah}, {s}^{\dagger}_{\rm av}, {s}^{\dagger}_{\rm bh}$ or
${s}^{\dagger}_{\rm bv}$ depending on the location. The
average waiting time for successful charging of all four
ensembles is approximately given by
$T=\frac{1}{rp}(\frac{1}{4}+\frac{1}{3}+\frac{1}{2}+1)=\frac{25}{12
rp}$. Thanks to the independent creation and storage, it
scales only like $1/p$. Once all ensembles are charged, the
four stored spin-wave modes are then \textit{partially}
converted back into a photonic excitations. This is done
using read pulses whose area is smaller than the standard
value of $\pi$, such that the state of the system is given
by $ (\alpha {a}'^{\dagger}_h+\beta {s}^{\dagger}_{\rm
ah})\otimes (\alpha {a}'^{\dagger}_v+\beta
{s}^{\dagger}_{\rm av})\otimes (\alpha
{b}'^{\dagger}_h+\beta {s}^{\dagger}_{\rm bh})\otimes
(\alpha {b}'^{\dagger}_v+\beta {s}^{\dagger}_{\rm
bv})|0\rangle $ with $|\alpha|^2+|\beta|^2=1.$ The primed
modes ${a}'^{\dagger}_h, {a}'^{\dagger}_v,$
$({b}'^{\dagger}_h, {b}'^{\dagger}_v)$ refer to the emitted
Anti-Stokes photons from memories located at $A_h$ and
$A_v$ ($B_h$ and $B_v$) respectively; $|0\rangle$ denotes
the empty state. The released Anti-Stokes photons are
combined at a central station where they are detected in
modes ${d}_{\pm}={a}'_{h}+{a}'_{v}\pm {b}'_{h}\mp {b}'_{v}$
and ${\tilde{d}}_\pm=\pm {a}'_{h} \mp
{a}'_{v}+{b}'_{h}+{b}'_{v}$, using the setup shown in Fig.
2. In the ideal case, a twofold coincident detection
between ${d}_+$ and ${\tilde{d}}_+$ projects the state of
the two remaining spin-wave modes non-destructively onto
\begin{equation}
|\Phi_{\rm{ab}}\rangle=1/\sqrt{2}({s}_{\rm{ah}}^{\dagger}{s}_{\rm{bh}}^{\dagger}+{s}_{\rm{av}}^{\dagger}{s}_{\rm{bv}}^{\dagger})|0\rangle.
\end{equation}
The stored atomic excitations can be reconverted into
photons as desired. In the proposed quantum repeater
protocol (cf. sec. \ref{Repeater}), one excitation (e.g.
the one in the $B$ ensembles) is reconverted into a photon
right away and used for entanglement generation. The other
excitation is reconverted later for entanglement swapping
or for the final use of the entanglement. Note that the
setup can also be used as a heralded source of single
photon pairs \cite{zhang,sliwa}.

Given an initial state where all four memories are charged,
the probability for a coincidence between ${d}_+$ and
${\tilde{d}}_+$ is given by $\frac{1}{2}\alpha^4 \beta^4.$
Since the twofold coincidences ${d}_+$-${\tilde{d}}_-$,
${d}_-$-${\tilde{d}}_+$, ${d}_-$-${\tilde{d}}_-$ combined
with the appropriate one-qubit transformation also collapse
the state of the atomic ensembles into
$|\Phi_{\rm{ab}}\rangle,$ the overall success probability
for the entangled pair preparation is given by
$P_s=2\alpha^4\beta^4.$

We now analyze the effect of non-unit detector efficiency
$\eta_d$ and memory recall efficiency $\eta_m$. The waiting
time for the memories to be charged is now
$T^\eta=T/\eta_d=\frac{25}{12 rp \eta_d}$. Furthermore, the
detectors can now give the expected coincidences when three
or four Anti-Stokes photons are released by the memories,
but only two are detected. In this case, the created state
contains additional terms including single spin-wave modes
and a vacuum component,
\begin{eqnarray}
\label{rho}
{\rho}_{\rm ab}^{s}&=&\nonumber
c_2^{s} |\Phi_{\rm{ab}}\rangle\langle \Phi_{\rm{ab}} | \\
&& \nonumber +c_1^{s} \Big( |{s}_{{\rm ah}}\rangle\langle {s}_{{\rm ah}}|
+|{s}_{{\rm av}}\rangle\langle {s}_{{\rm av}}|
+|{s}_{{\rm bh}}\rangle\langle {s}_{{\rm bh}}|
+|{s}_{{\rm bv}}\rangle\langle {s}_{{\rm bv}}|\Big)\\
&&+c_0^{s} |0\rangle\langle0|;
\end{eqnarray}
where $c_2^{s}=2\alpha^4\beta^4\eta^2/P_s^{\eta},$
$c_1^{s}= \alpha^6 \beta^2\eta^2(1-\eta)/P_s^{\eta}$ and
$c_0^{s}=2\alpha^8(1-\eta)^2\eta^2/P_s^{\eta}$. Here
$\eta=\eta_m\eta_d$ is the product of the memory recall
efficiency and the (photon-number resolving) detector
efficiency, and we have introduced a superscript $s$ for
``source''. The probability for the successful preparation
of this mixed state is $P_s^{\eta} = 2 \eta^2
\alpha^4(1-\alpha^2\eta)^2.$ The fidelity of the
conditionally prepared state is equal to the two-photon
component $c_2^s={\beta^4}/{(1-\alpha^2\eta)^2}.$ As can be
seen from the two previous equations, there is a tradeoff
on the readout coefficients $\alpha, \beta$. The creation
of an entangled state with a high fidelity favors $\alpha
\approx 0$, whereas a high success probability favors
$\alpha \approx \beta \approx 1/\sqrt{2}.$

\section{Repeater Protocol}
\label{Repeater}

 We now include our source of heralded pairs within a quantum repeater protocol following
 Ref. \cite{Chen07}. Fig. 3A shows how entanglement between
 two remote sources (denoted AB and CD) is
 created by combining two Anti-Stokes photons at a central
 station, where one photon is released from the B ensembles
 and the other from the C ensembles, and performing a
 projective measurement into the modes
 ${D}_\pm^{bc}={b}'_{\rm{h}}\pm {c}'_{\rm{v}}$ and
 ${D}_{\pm}^{cb}={c}'_{\rm{h}}\pm {b}'_{\rm{v}}$ using the same combination of
 linear optical elements as in Refs. \cite{Jiang07, Chen07, Zhao07}.
 The twofold coincident detection ${D}_+^{bc}$-${D}_+^{cb}$
 (${D}_+^{bc}$-${D}_-^{cb}$, ${D}_-^{bc}$-${D}_+^{cb}$, or
 ${D}_-^{bc}$-${D}_-^{cb}$ combined with the appropriate
one-qubit operations) collapses the two remaining full
memories into $|\Phi_{\rm ad}\rangle.$ Due to
imperfections, the distributed state $\rho_{\rm ad}^{0}$
includes vacuum and single spin-wave modes. One can show
that their weights $c_2^0, c_1^0, c_0^0$ are unchanged
compared with the weights of the source state $\rho_{\rm
ab}^s,$ because $c_2^0=\frac{(c_2^s)^2}{(c_2^s+2
c_1^s)^2}=c_2^s$, $c_1^0=\frac{c_1^s c_2^s}{(c_2^s+2
c_1^s)^2}=c_1^s$ and $c_0^0=\frac{4 (c_1^s)^2}{(c_2^s+2
c_1^s)^2}=c_0^s$. (The condition for having a stationary
state is $c_0 c_2=4 (c_1)^2$, which is fulfilled by
$c_2^s,c_1^s,c_0^s$.) The success probability for the
entanglement creation is given by
$P_0=2\eta^2\eta_t^2\left(c_2^s/2+c_1^s \right)^2.$ Here
$\eta_t$ is the fiber transmission for each photon.

Fig. 3B shows how, using the same combination of linear
optical elements and detectors, one can perform successive
entanglement swapping operations, such that the state
$\rho_{\rm az}^{n}$ is distributed between the distant
locations A and Z after $n$ swapping operations. In analogy
to above, one can show that the distributed state
$\rho_{\rm az}^{n}$ includes vacuum and single spin-wave
components with unchanged weights with respect to the
initial ones, i.e. $c_2^{n}=c_2^{s}, $ $c_1^{n}=c_1^s$ and
$c_0^{n}=c_0^s.$ From the expression of $P_0$ and keeping
in mind that the entanglement swapping operations are
performed locally such that there are no transmission
losses, one deduces the success probability for the $i$-th
swapping, $P_i=2\eta^2\left(c_2^s/2+c_1^s \right)^2.$ The
two-spin-wave component of the distributed mixed state
$|\Phi_{\rm az}\rangle$ is finally post-selected
with the probability $P_{\rm pr}=c_2^s\eta^2.$\\

\begin{figure}
{\includegraphics[scale=0.38]{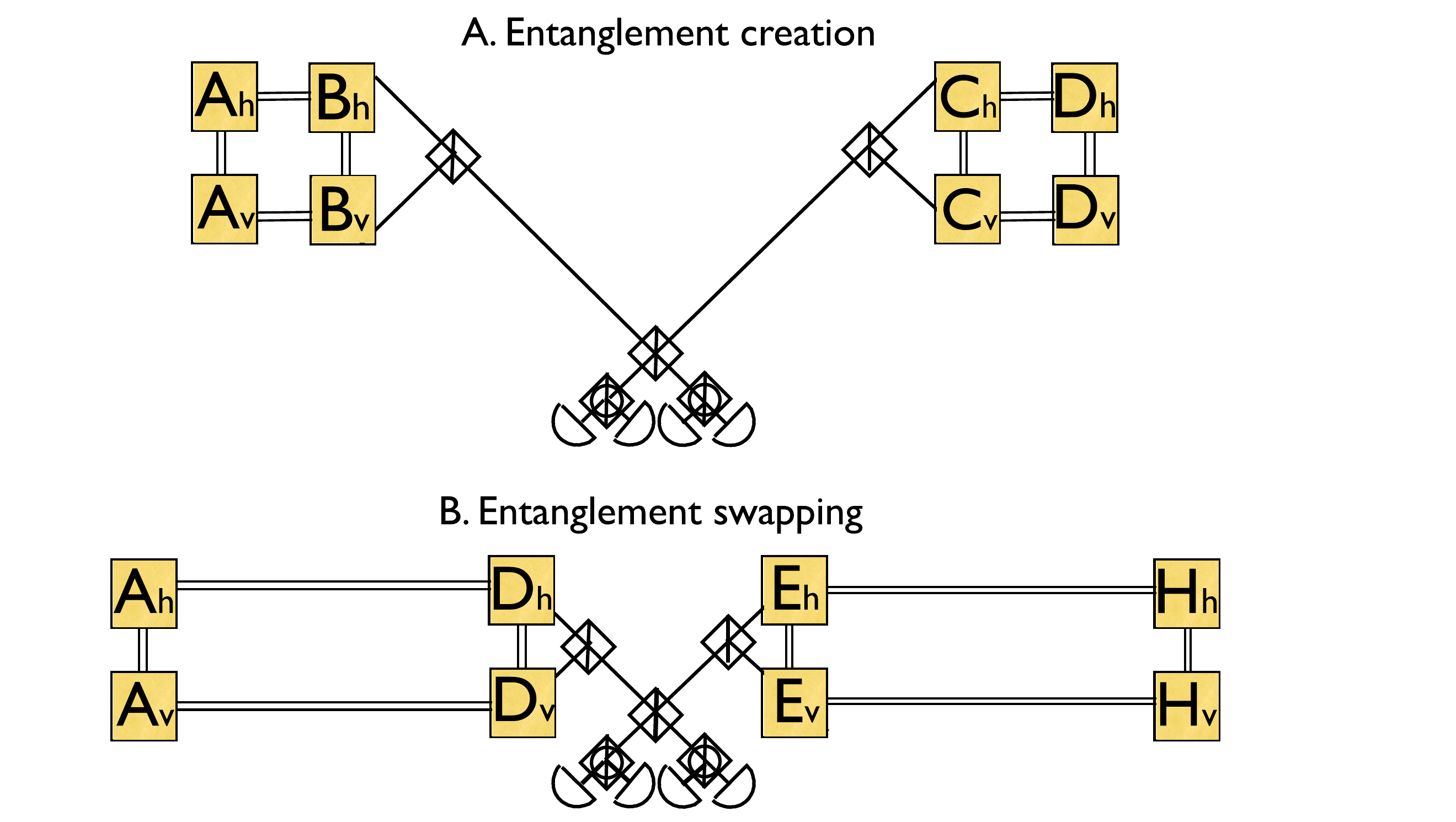} \caption{(Color
online) (A) Long-distance entanglement creation using two
four-ensemble sources as shown in Fig. 2. The A and D
ensembles are entangled by the detection of two photons
emitted from the B and C ensembles, using the same setup as
in Refs. \cite{Jiang07, Chen07, Zhao07}. Note that the AB
source is separated from the CD source by a long distance.
(B) Entanglement swapping. The same set of linear optical
elements allows one to entangle the A and H ensembles
belonging to two adjacent elementary links. Note that the D
and E ensembles are at the same location.}\label{fig3}}
\end{figure}

The time required for a successful distribution of an entangled state
$|\Phi_{\rm az}\rangle$ is approximately given by
\cite{Simon07}
\begin{equation}
\label{total_time} T_{\rm
tot}=\left(\frac{3}{2}\right)^{n}\frac{L_0}{c}
\frac{1}{P_0P_1...P_n P_{\rm pr}},
\end{equation}
where $L_0=L/2^n$ is the length of an elementary link, $L$
is the total distance and $n$ is the nesting level of the
repeater. Taking into account the expressions of $P_0,$
$P_i$ (with $i \geq 1$) and $P_{\rm pr}$, one can rewrite
$T_{\rm tot}$ as
\begin{equation}
T_{\rm tot}=2\times3^{n} \times \frac{L_0}{c}
\frac{(1-\alpha^2\eta)^{2(n+2)}}{\eta_{t}^2\eta^{2(n+2)}\beta^{4(n+2)}}.
\end{equation}
Here $\eta_t=e^{-L_0/(2 L_{att})}$ is the fiber
transmission, with the attenuation length $L_{att}$. In our
numerical examples we use $L_{att}= 22$ km, corresponding
to losses of 0.2 dB/km, which are currently achievable at a
wavelength of 1.5 $\mu$m \cite{crypto-review}; $c = 2
\times 10^8$ m/s is the photon velocity in the fiber.

For these formulas to be strictly valid, the source
preparation time has to be negligible compared to the
communication time, i.e. in our case $T_s=\frac{3 T^\eta}{2
P_s^\eta} \ll \frac{L_0}{c}$. Otherwise one simply has to
replace $\frac{L_0}{c}$ by $\frac{L_0}{c}+T_s$, cf. below.

We now consider the role of errors due to the creation of
two excitations in a single memory. Note that in the local
entanglement generation process of Fig. 2 a large part of
such multi-photon events will be detected because both
Stokes and Anti-Stokes photons are detected locally and
thus potentially with high efficiency. We find by explicit
calculation that the fidelity of the distributed state at
the first order in $p$ after $n=4$ swapping levels
(neglecting other errors) is given by
\begin{equation}
\nonumber F \approx
{1-\left[\left(418-260\eta)+(47-205\eta\right)\alpha^2\right](1-\eta_d)p}.
\end{equation}
If one wants a fidelity of the final state $F=0.9$, one can
choose e.g. $\alpha^2=0.2$ and $p=6\times10^{-3}$. This is
the value of $\alpha$ used in Fig. 1. For these values,
equality between the source preparation time $T_s$ and the
communication time $L_0/c$ (e.g. for $L=1000$ km and 16
links) is reached for a basic repetition rate $r$ of order
60 MHz, cf. below.

\section{Implementation}
\label{Implementation}

In this section we discuss potential experimental
implementations of the proposed protocol. There has
recently been impressive progress on the efficiency of
conversion from atomic excitation to photon, which sets the
fundamental limit for the memory efficiency. Values as high
as 84 percent have been achieved with a cavity setup
\cite{simon-vuletic}.

Current DLCZ-type experimental setups with atomic gases
\cite{dlcz_expts,chou,yachen} are very well suited for
demonstrating the proposed ideas. Current repetition rates
$r$ in DLCZ-type experiments are of order a few MHz. To
fully exploit the potential of the proposed protocol, the
rates have to be increased, cf. above. Rates of tens of
MHz, which could already bring the overall entanglement
generation times to within a factor of 2 or 3 of the values
given in Fig. 1, are compatible with typical atomic
lifetimes. With atomic gases, further improvements in $r$
could be achieved using the Purcell effect in high-finesse
cavities to reduce the atomic lifetimes.

Ref. \cite{Simon07} pointed out that the combination of a
photon pair source and of a quantum memory which stores one
of the photons is equivalent to a DLCZ-type atomic
ensemble, which emits a photon that is correlated with an
atomic excitation. This approach may make it possible to
achieve even higher values of $r$, using e.g. photon pair
sources based on parametric down-conversion and solid-state
quantum memories based on controlled reversible
inhomogeneous broadening \cite{CRIB}. Solid-state atomic
ensembles, e.g. rare-earth ion doped crystals, furthermore
hold the promise of allowing very long storage times (which
are essential for quantum repeaters), since the storage
time is no longer limited by atomic motion, while the
intrinsic atomic coherence times can be very high. For
example, hyperfine coherence times as long as 30 s have
been demonstrated in Pr:Y$_2$SiO$_5$ \cite{hyperfine}. The
best efficiency published so far for a CRIB memory (in the
same material) is 15 \% \cite{hetet}, but experiments are
progressing quite rapidly. This approach furthermore holds
the promise of allowing temporal multiplexing
\cite{Simon07}, leading to a potential further improvement
in the entanglement creation rate, provided that multi-mode
memories with the required characteristics can be realized.
The main requirements are sufficient optical depth and
sufficient memory bandwidth. Other forms of multiplexing
could also be possible and might allow to relax the
requirements on the memory storage times \cite{collins}.
Ideally the memories in the described protocol should
operate at the optimal wavelength for telecom fibers, i.e.
at 1.5 $\mu$m. This may be possible with Erbium-doped
crystals \cite{staudt}. Alternatively, wavelength
conversion techniques could be employed \cite{tanzilli}.

Good photon detectors with photon number resolution are
also essential. Superconducting transition-edge sensor
detectors can already resolve telecom-wavelength photons of
4 ns duration at a repetition rate of 50 kHz, with an
efficiency of 0.88 and negligible noise \cite{Rosenberg05}.
In the long run, NbN detectors are promising for achieving
higher rates. The detection of 100 ps photons with 100 MHz
rate has been reported in ref. \cite{Rosfjord} with an
efficiency of 0.56 and a noise smaller than 10/s.

Our results show the great interest for quantum repeaters
of locally generating entangled pairs of excitations with
high fidelity. This could also be achieved for physical
systems other than atomic ensembles. Promising approaches
include the creation of atom-photon entanglement
\cite{atom-photon} and entangled photon pair sources based
on quantum dots \cite{qdots}, which could be combined with
quantum memories.

\section{Conclusions}
We started this work with a quantitative comparison of
different quantum repeater protocols using only atomic
ensembles and linear optics. This comparison showed that
protocols based on the local generation of high-fidelity
entangled pairs of atomic excitations make it possible to
combine robustness with respect to phase fluctuations and
good entanglement distribution rates. We then proposed a
new approach for local entanglement generation based on
partial memory readout. Together with the use of two-photon
detections for long-distance entanglement generation and
for entanglement swapping, this approach leads to a
repeater protocol that, as far as we know, achieves the
highest entanglement distribution rate with the given
ingredients. First demonstration experiments should be
possible with atomic gases. The protocol could reach its
full potential combining fast photon pair sources such as
parametric down-conversion and solid-state quantum
memories.

We thank R. Dubessy  for useful discussions. This work was
supported by the EU via the Integrated Project {\it Qubit
Applications (QAP)} and a Marie Curie Excellence Grant, by
the Swiss NCCR {\it Quantum Photonics}, and by the Chinese
National Fundamental Research Program (No. 2006CB921900).


\begin{thebibliography}{100}
\bibitem{partial-sps} The partial readout approach
described in Sec. \ref{Pairs} can be applied to the scheme
of Ref. \cite{Sangouard07}, making it possible for one
ensemble to serve as single-photon source and memory at the
same time.
\end{thebibliography}
\end{document}